\documentclass[pra,twocolumn,superscriptaddress]{revtex4}

\usepackage{amsfonts}
\usepackage{amsmath}
\usepackage{bbm}
\usepackage{graphicx}
\usepackage{bbold}
\usepackage{amssymb}
\usepackage{color}
\usepackage{subfigure}
\usepackage{mathtools}

\begin{document}
\title{Simulation of time-crystal-like behavior for a few boson \\chiral soliton model in a ring}

\author{Patrik \"Ohberg}
\affiliation{SUPA, Institute of Photonics and Quantum Sciences, Heriot-Watt University, Edinburgh EH14 4AS, United Kingdom}
\author{Ewan M. Wright}
\affiliation{SUPA, Institute of Photonics and Quantum Sciences, Heriot-Watt University, Edinburgh EH14 4AS, United Kingdom}
\affiliation{James C. Wyant College of Optical Sciences, University of Arizona, Tucson, Arizona 85721, USA}

\date{\today}
\tolerance = 1000

\begin{abstract}
We present numerical simulations for a chiral soliton model with $N=2,3$ bosons in a ring, this being a few-particle version of our previous mean-field model for a quantum time crystal \cite{OW19}.  Following Syrwid, Kosior, and Sacha (SKS), the notion is that a precise position measurement of one particle can lead to spontaneous formation of a bright soliton that in a time crystal should rotate intact for at least a few revolutions around the ring \cite{SKS_EPL}.  In their work SKS find spontaneous formation of a soliton due to the position measurement, but quantum fluctuations cause the soliton to subsequently decay before it has a chance to perform even one revolution of the ring.  Based on this they conclude that time crystal dynamics are impossible for Wilczek's model of a bright soliton in a ring. In contrast, for our few boson chiral soliton model, and allowing for imprecise (weak) measurements of the particle position, we show that  time-crystal-like behavior is possible allowing for several revolutions of the spontaneously formed soliton around the ring.
\end{abstract}
\maketitle

\section{Introduction}

Since Wilczek's seminal paper initiated the field of quantum time crystals in 2012 \cite{Wil12}, it has attracted intense interest, both fundamental and applied.  His model comprised $N$ bosons with attractive interactions in a ring that was pierced by a flux tube, the key idea being that this model could yield a ground state composed of a soliton that was rotating around the ring.  Although this model was subsequently shown to always yield a non-rotating ground state \cite{Bruno}, the time crystal idea had taken root.  Now there is active research in both continuous and discrete time crystals and for a variety of platforms including optics, ultracold atoms, metamaterials, spin systems, and both closed and open systems.  For recent reviews of the area see \cite{SacZak18,Sacha20,ZalLukMon23}.

A genuine time crystal involves a Hamiltonian system that exhibits sustained periodic motion even in its lowest-energy state. To the best of our knowledge there is only one example of such a system, that involves long range interactions, but with no physical implementation to date \cite{KozKyr19}.  The issue undermining Wilczek's initial proposal for a genuine time crystal was nicely illustrated in a paper by Syrwid, Kosior, and Sacha (SKS) in which they simulated N-bosons in a ring that is pierced by a flux tube, and which was initiated from the N-particle ground state which shows no signatures of rotation or localization \cite{SKS_EPL}.  What they showed was that precise position measurement of one particle could indeed lead to the spontaneous formation of an $(N-1)$-particle soliton that is rotating, and they suggested that if this initial soliton could survive several revolutions around the ring then one could at least speak of time-crystal-like behavior for the Wilczek model.  However, SKS found that quantum fluctuations cause this initial soliton to subsequently decay before it has a chance to perform even one revolution of the ring: They presented results for particle numbers in the range $N=10-60$, and the lifetime $t_c$ of the soliton was found to increase with $N$.  We observed the same behavior for a non-chiral version of our model with $N=3$, both for precise and imprecise position measurements (with an uncertainty in the position measurement),  the lifetime being longer for the imprecise measurement.  The conclusion of SKS remains intact that Wilczek's model can realize neither a genuine time crystal nor time-crystal-like behavior \cite{SKS_EPL}.

In 2019 we proposed a mean-field chiral soliton model for a quantum time crystal \cite{OW19}.  This paper was followed by discussion in the literature with SKS \cite{SKS_comment}-\cite{OW_lack}, culminating in their paper cited above \cite{SKS_EPL}.  The goal of the present paper is to examine a few boson $(N=2,3)$ limit of our previous chiral soliton model to numerically assess whether time-crystal-like behavior is possible in our model.  In particular, we find the $N$-boson ground state, perform a position measurement of one particle, and see if spontaneous formation of the resulting $(N-1)$-boson soliton can persist over several revolutions of the ring.  We find that for an imprecise or weak position measurement  time-crystal-like behavior is possible in our chiral soliton model, whereas for a precise measurement quantum fluctuations cause the soliton to decay in accordance with SKS \cite{SKS_EPL}.  We therefore argue that our chiral model can display time-crystal-like behavior when weak position measurements are employed.

The chiral model we discuss here has recently been experimentally realised in the meanfield limit by Fr\"olian et al \cite{frolian} and theoretically studied in \cite{edmonds,valenti,chisholm}. This chiral scenario with its density dependent nonlinear gauge potential opens up a number of intriguing possibilities to study topological gauge theories which can emerge as the low-energy limit of strongly correlated systems such as for instance the Chern-Simons theory of fractional quantum Hall phenomena. The experimental realisation of  the chiral nonlinear situation \cite{frolian} utilises a weakly interacting Bose-Einstein condensate where two internal atomic states are coupled by a laser. The experimental protocol relies on the fact that scattering lengths can be different for the two bare internal states. This in turn results in the corresponding atom-photon dressed states having momentum dependent effective collisional interaction strengths, which can be mapped onto a density dependent gauge potential.

The remainder of this paper is organized as follows: Section \ref{CSM} describes the few boson version of our chiral soliton model along with a description of the parameters involved.  The case of $N=2$ particles is covered in Sec. \ref{2PC} and shows the ideas involved in a transparent manner.  Section \ref{3PC} contains the main results and covers the three particle case that demonstrates that our few boson chiral soliton model can display time-crystal-like behavior in which a two-boson soliton state \cite{Jackiw} can execute several revolutions of the ring, and the soliton lifetime is explored as a function of the position uncertainty of the position measurement. 
Summary and conclusions are given in Sec. \ref{SC}.

\section{Chiral soliton model}\label{CSM}

\subsection{N-particle Schr\"odinger equation}
We consider a system of $N$ scalar bosons of mass $m$ located circumferentially in a ring of radius $R$.  The Schr\"odinger equation for the N-particle wavefunction $\psi(\theta_1,\theta_2,\ldots \theta_N,t)$ is in scaled units \cite{PriValWri23}
\begin{eqnarray}
i{\partial\Psi\over\partial t} &=& \sum_{j=1}^N  \left ( -i{\partial\over\partial \theta_j}-A(\theta_1,\theta_2,\ldots \theta_N) \right )^2 \Psi  \nonumber \\ &&+{g\over 2} \rho(\theta_1,\theta_2,\ldots \theta_N) \Psi .
\end{eqnarray}
Here $\theta_j=[-\pi,\pi], j=1,2,\ldots N$ are angular coordinates around the ring, and time is in units of $2mR^2/\hbar$.  The scaled gauge potential is given by
\begin{equation}\label{Apot}
A(\theta_1,\theta_2,\ldots \theta_N) = A^{(0)} + \kappa \rho(\theta_1,\theta_2,\ldots \theta_N), 
\end{equation}
where $A^{(0)}$ represents the single-particle contribution to the gauge potential, and $\rho(\theta_1,\theta_2,\ldots \theta_N) $ describes the density-dependent contribution to the gauge potential.  The dimensionless parameter $\kappa$ characterizes the strength of the density-dependent chiral gauge potential, and the parameter $g$ controls the density-dependent and non-chiral contribution to the system energy.  The density-dependent gauge potential is built from the combinations of the two-particle interactions between the composite bosonic particles, the two-particle interaction being denoted $\eta(x)$, with $x$ the separation between the two particles.  In particular, we set
\begin{equation}
\rho(\theta_1,\theta_2,\ldots \theta_N) = \sum_{i=1}^N \sum_{j>i}^N \eta(\theta_i-\theta_j),
\end{equation}
where $\rho(\theta_1,\theta_2,\ldots \theta_N)$ is symmetric under particle exchange by virtue of the symmetry $\eta(x)=\eta(-x)$.  Since the density and corresponding density-dependent gauge potential are symmetric under particle exchange, which breaks Galilean invariance, we expect chiral dynamics to appear in this model.

\subsection{Model parameters}

In our previous mean-field model the chiral soliton model was derived starting from the spinor Schr\"odinger equation for a system of two-level bosonic atoms trapped in a ring that are dipole coupled using a laser beam carrying orbital angular momentum (OAM) characterized by the winding number $\ell$.  By preparing the atoms in the appropriate dressed state a scalar field analysis may be obtained in which the atoms are subject to a vector potential of the form in Eq. (\ref{Apot}) with $A^{(0)}={\ell\over 2}$.  Moreover, the density-dependent gauge potential results from a collision-induced detuning, with the consequence that the dimensionless parameter $\kappa\propto \ell$.  This means that if the winding number of the laser beam $\ell=0$, then $\kappa=0$ and there will be no chiral dynamics and no possibility for time-crystal-like behavior.  The term proportional to the dimensionless parameter $g$ describes energy-shifts due to many-body interactions. 

For numerical purposes it is impractical to consider s-wave many-body interactions with $\eta(x)=\delta(x)$.  Instead we here use the non-local model \cite{PriValWri23}
\begin{equation}
\eta(x) = Q\cos^{2q}(x/2) , \quad q=1,2,3,\ldots ,
\end{equation}
where $Q$ is a normalization constant such that $\int_{-\pi}^\pi \eta(x) dx = 1$. This allows for numerical simulations to be performed in conjunction with checking that the results obtained are not particularly sensitive to the choice of $q$.  In the simulations presented we set $q=50$ but increasing this to $q=100$ made little difference, meaning that the non-locality is not a key ingredient for the results we obtain.  In this sense our model is local for all practical purposes.

\section{Two-particle case}\label{2PC}

As a first example we consider the case of two bosons in a ring.  This example is over-simplified in that if one particle is measured this leaves only one particle on the ring, so that the subsequent dynamics is linear.  This example is nonetheless quite transparent and illustrates the ideas involved.

\subsection{Two-particle Schr\"odinger equation}

The density for the two bosonic particles may be written as
\begin{eqnarray}
\rho(\theta_1,\theta_2)&\equiv & \eta(\theta_1-\theta_2), 
\end{eqnarray}
and the Schr\"odinger equation for $\Psi(\theta_1,\theta_2,t)$ then becomes \cite{PriValWri23}
\begin{eqnarray}
i{\partial\Psi\over\partial t} &=& \sum_{j=1}^2  \left (-i{\partial\over \partial \theta_j} - {\ell\over 2} -\kappa\eta(\theta_1-\theta_2) \right )^2 \Psi \nonumber \\ && + {g\over 2} \eta(\theta_1-\theta_2) \Psi .
\end{eqnarray}
Changing to center-of-mass (COM) and relative coordinates using $s=(\theta_1+\theta_2)/2, x=(\theta_1-\theta_2)$, the transformed Schr\"odinger equation for $\Psi(x,s,t)$ becomes 
\begin{eqnarray}\label{PsiEq}
i{\partial\Psi\over\partial t} &=& \bigg[
-2 {\partial^2\over \partial x^2} - {1\over 2}  {\partial^2\over \partial s^2} + 2i\left ( {\ell\over 2} + \kappa\eta(x) \right ) {\partial\over\partial s} \nonumber \\ &&
+ 2 \left ( {\ell\over 2} + \kappa\eta(x) \right )^2 + {g\over 2} \eta(x) \bigg] \Psi .
\end{eqnarray}

\subsection{Two-particle ground state}

We seek a two-particle stationary state using the ansatz \cite{PriValWri23}
\begin{equation}\label{SS}
\Psi(x,s,t) = e^{-i\varepsilon t + ips}\varphi(x),
\end{equation}
with $\varepsilon$ the scaled energy of the two-particle system, and $p$ an integer that is the winding number associated with the center-of-mass OAM of the two particles.  Substituting this ansatz into Eq. (\ref{PsiEq}) yields
\begin{equation}\label{phiEq}
\left [
-{d^2\over dx^2}+ \left ( {(p-\ell)\over 2} - \kappa\eta(x) \right )^2 + {g\over 2} \eta(x) \right ] \varphi(x) = {\varepsilon\over 2} \varphi(x) ,
\end{equation}
where $\varepsilon$ is the scaled energy of the two-particle system.

\begin{figure}[h!]
\includegraphics*[width=9cm]{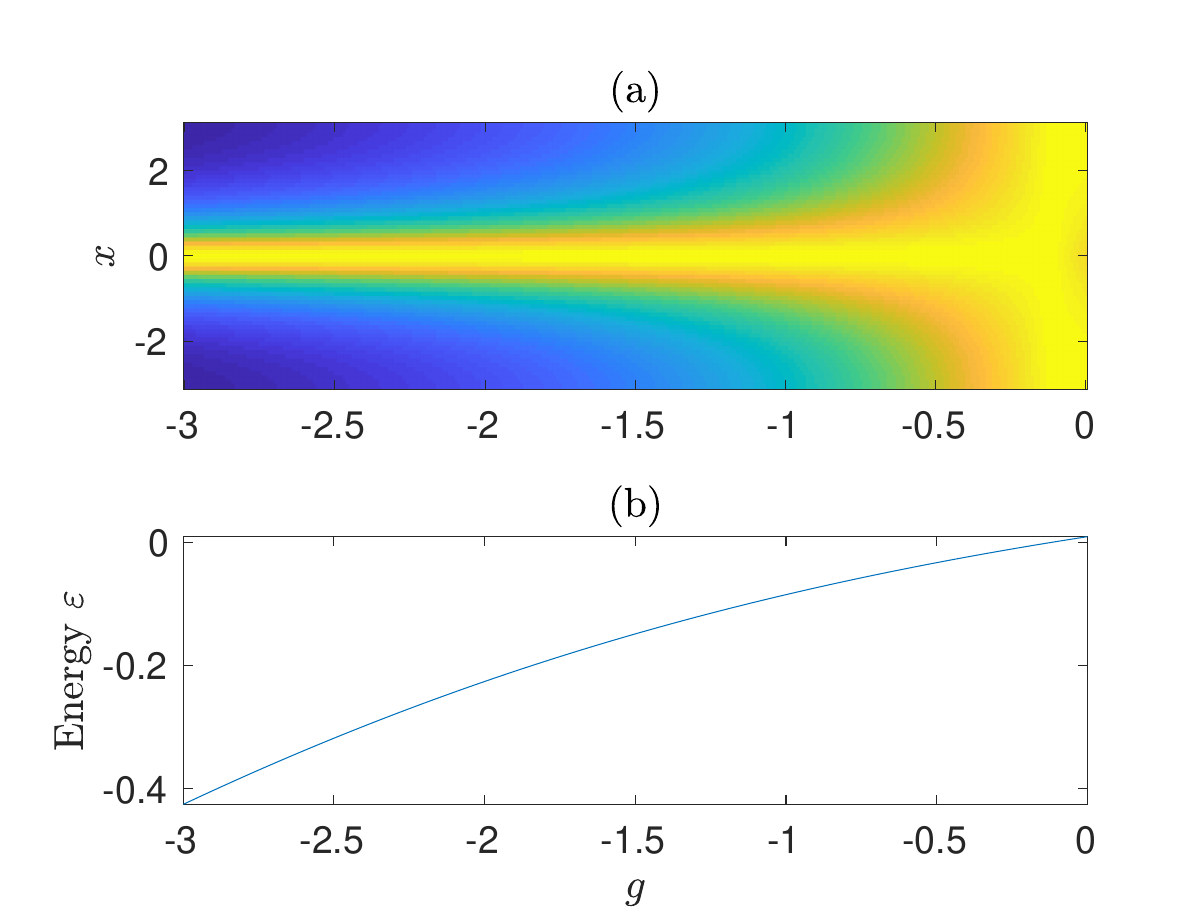}
\caption{(a) Ground state profile $|\varphi(x)|^2$ versus $x$ and $g$, and (b) scaled energy $\varepsilon$ versus $g$.  Parameters used are $N=2, \ell=2$, $\kappa=0.2$, and $q=50$.}\label{Fig1}
\end{figure}

\begin{figure}[h!]
\centering
\includegraphics*[width=8cm]{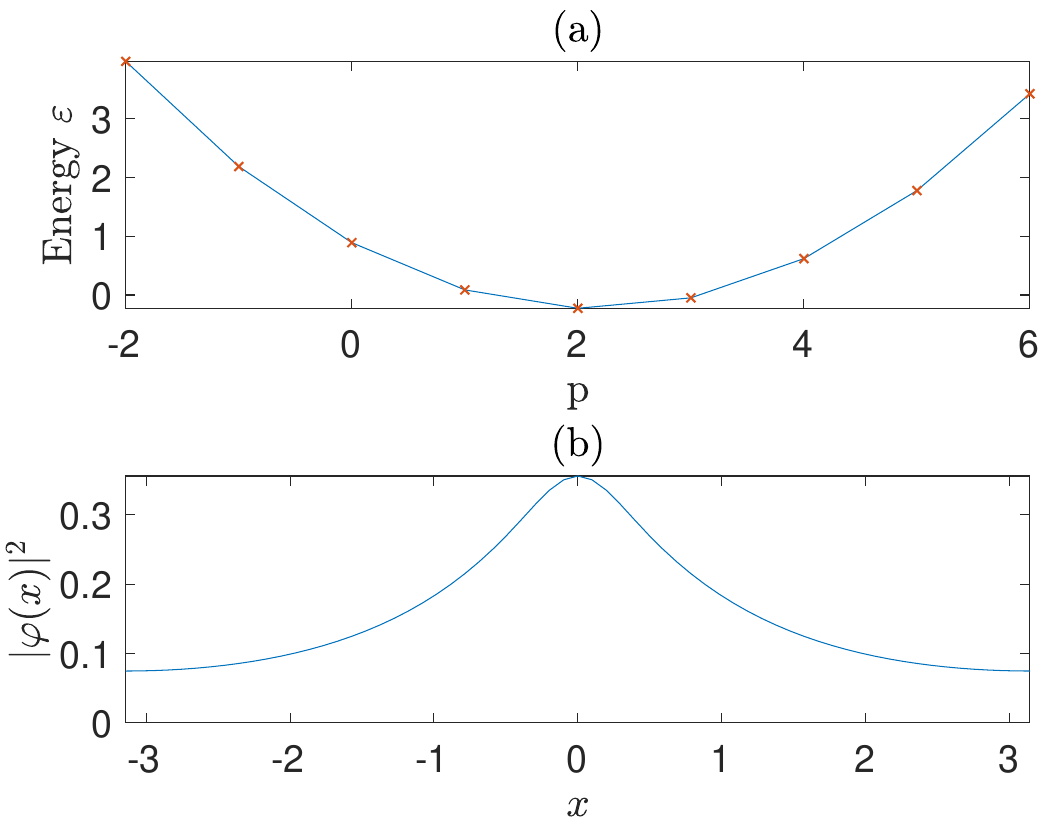}
\caption{(a) Scaled energy $\varepsilon$ versus COM winding number $p$, and (b) ground state profile $|\varphi(x)|^2$ versus $x$.  Parameters are the same as Fig. \ref{Fig1} along with $g=-2$.}\label{Fig2}
\end{figure}

We have solved Eq. (\ref{phiEq}) numerically for a range of parameters using the imaginary time method \cite{LehToiElo2007} applied to Eq. (\ref{PsiEq}), in particular to obtain the ground state $\varphi(x)$ and associated scaled energy $\varepsilon$.  Figure \ref{Fig1} shows numerical results for parameters $\ell=2,\kappa=0.2, q=50$.  Specifically, Fig. \ref{Fig1}(a) shows a color coded plot of the ground state profile $|\varphi(x)|^2$ versus $x$ for a variety of $g<0$ along the horizontal axis, and plot (b) shows the scaled energy $\varepsilon$ over the same range of $g$.  The ground state profile $|\varphi(x)|^2$ is normalized to unity on-axis so that the profile can be discerned for each value of $g$.  What is evident is that for $g > -0.5$ the ground state is uniform to a high degree, whereas for $g < -0.5$ the ground state profile is localized in terms of the relative coordinate $x$.

In the remainder of this section we choose $g=-2$ as an illustrative example, though similar results could be obtained for different combinations of parameters. Figure \ref{Fig2} shows some more detail of the solution for this choice.  In particular, Fig. \ref{Fig2}(a) shows the scaled energy versus $p$ and reveals that the ground state occurs for $p=\ell=2$, and (b) shows a line plot of the ground state profile $|\varphi(x)|^2$ (properly normalized) versus $x$ for $p=2$.  Furthermore, the width (full-width at half-maximum) of the ground state profile in Fig. \ref{Fig2}(b) with respect the the relative coordinate $x$ of the two particles is $w \sim 1.6$, the parameter $w$ characterizing the expected width of any localized two-particle solutions that may arise. 

\begin{figure}[h!]
\centering
\includegraphics*[width=5cm]{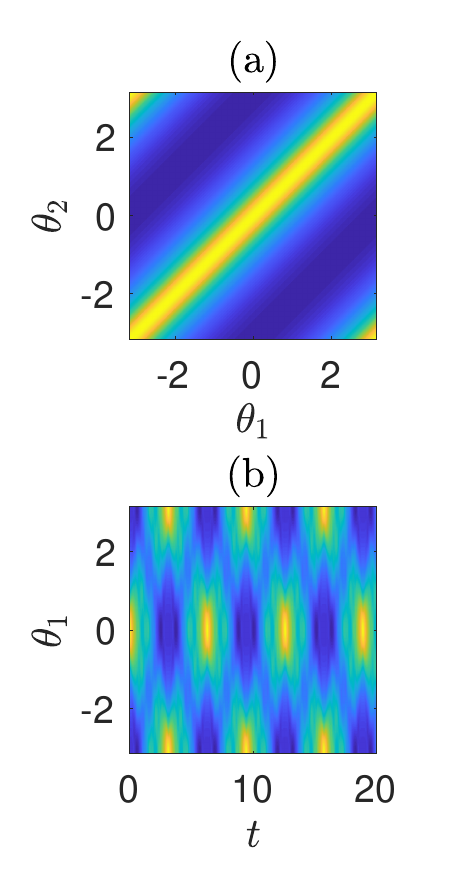}
\caption{(a) Reconstructed ground state probability density $|\Psi(\theta_1,\theta_2)|^2$, and (b) color coded plot of the probability density $p(\theta_1,t)$ for a measurement with $n=50$ that localizes particle $2$ with an uncertainty $\Delta\theta\sim 0.34$.  This example does not display time-crystal-like behavior.  Parameters are the same as as in Fig. \ref{Fig1} along with $g=-2$.}\label{Fig3}
\end{figure}

\begin{figure}[h!]
\centering
\includegraphics*[width=5cm]{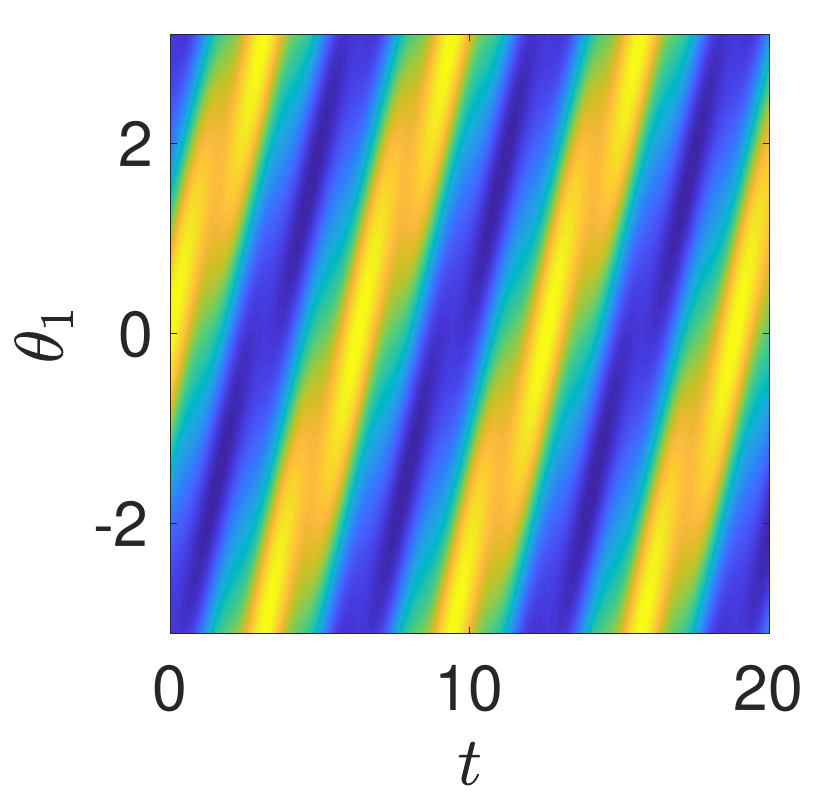}
\caption{Color coded plot of the probability density $p(\theta_1,t)$ for an imprecise measurement with $n=1$ and an uncertainty $\Delta\theta\sim 2.3$ that dominantly couples the remaining particle after the measurement of particle 2 to a non-dispersing solution for the ring. This example does display time-crystal-like behaviour.  Parameters are the same as as in Fig. \ref{Fig1} along with $g=-2$.}\label{Fig4}
\end{figure}

\subsection{Time-crystal-like dynamics from measurement}

To proceed we first recognize that the ground state represented by the probability density in Fig. \ref{Fig2}(b) does not correspond to a localized state on the ring since $|\varphi(x)|^2$ depends only on the relative coordinate $x$, and there is no information concerning the COM position $s$ of the particle. On the other hand, since the ground state has a winding number $p\ne 0$ it does represent a persistent flow \cite{RyuAndCla07}. We can probe further and reconstruct the ground state two-body wave function $\Psi(\theta_1,\theta_2)$ (at $t=0$), where $|\Psi(\theta_1,\theta_2)|^2$ is shown in Fig. \ref{Fig3}(a).  From this it is straightforward to see that the probability density of finding the particle at $\theta_1$ for any value of $\theta_2$,
\begin{equation}
    P(\theta_1) = \int_{-\pi}^\pi~d\theta_2 |\Psi(\theta_1,\theta_2)|^2 = {1\over 2\pi},
\end{equation}
is a flat distribution, once again verifying that for the ground state the solution does not represent a localized solution that is pinned to a given COM position.  In their work SKS \cite{SKS_EPL} argued that a precise position measurement of one particle could lead initially to spontaneous formation of a localized solution, and we pursue and expand that concept here for our chiral soliton model.  In particular, we assume that the position of particle $2$ is measured yielding a value $\theta_2^{(0)}$, but allow for the measurement to be imprecise, so that the wave function for the remaining particle may be written as
\begin{equation}\label{psi_init}
    \psi(\theta_1,t=0)=\Psi(\theta_1,\theta_2^{(0)}) = \int_{-\pi}^\pi d\theta_2~G(\theta_2-\theta_2^{(0)})\Psi(\theta_1,\theta_2) ,
\end{equation}
where the function $G(\theta_2-\theta_2^{(0)})$ reflects the uncertainty of the position measurement of particle $2$, and is a distribution which obeys the periodicity of the ring, and with an adjustable width $\Delta\theta$ (full-width at half-maximum of $|G(\theta)|^2$).  Here for illustration we employ the following distribution
\begin{equation}\label{G}
G(\theta-\theta^{(0)}) = {\cal N}\cos^{2n}\left ( {\theta-\theta^{(0)}\over 2}\right ) , \quad n=1,2,3,\ldots 
\end{equation}
with ${\cal N}$ a normalization constant.  In the limit $n\gg 1$ this distribution approaches a delta-function which corresponds to the precise measurement model adopted in the paper of SKS, whereas for $n=1$ the measurement uncertainty is around $\Delta\theta\sim 2.3$.  After the measurement, precise or imprecise,  the wave function $\psi(\theta_1,t=0)$ (suitably normalized) for the remaining particle is given by Eq. (\ref{psi_init}), and subsequent time development is governed by the single-particle Schr\"odinger equation
\begin{equation}\label{EqNeq1}
i{\partial\psi\over\partial t} = \left ( -i{\partial\over\partial \theta_1}-{\ell\over 2} \right )^2 \psi .
\end{equation}
Taking into account the $e^{ip\theta_1/2}$ angular variation associated with the ground state solution (\ref{SS}), the general solution of this equation is
\begin{equation}
\psi(\theta_1,t) = {e^{ip\theta_1/2}\over \sqrt{2\pi}} \sum_{j=-\infty}^\infty c_j e^{-ij^2 t},
\end{equation}
with expansion coefficients $c_j$.  This general solution reveals the fact that the single-particle quantum dynamics must be periodic in time with period $2\pi$.

Figure \ref{Fig3}(b) shows a simulation of the single-particle probability density $p(\theta_1,t)=|\psi(\theta_1,t)|^2$ versus $t$ and $\theta_1$ according to Eq. (\ref{EqNeq1}) with $\theta_2^{(0)}=0$ without loss of generality, and $n=50$ so that $\Delta\theta\sim 0.34$. We chose $n=50$ as this approaches a precise measurement as closely as possible within our non-local model with $q=50$ in Eq. (\ref{G}).  In this case the uncertainty is much less than the expected width $w\sim 1.6$ for a two-particle localized state, so the measurement is expected to be very disruptive.  We note that, as expected, the probability density varies periodically in time with period $2\pi$.  This temporal periodicity might suggest time-crystal-like behavior but this is not the case since the initial probability density profile $p(\theta_1,t=0)$ becomes rapidly distorted as time increases, not simply translated in time as expected for a time crystal.  This is in perfect keeping with the findings of SKS that too precise a position measurement of one particle can produce a localized state initially but this will become rapidly dispersed before the initial localized solution can exhibit even a single rotation around the ring.

For time-crystal-like behavior what is required is that the center-of-mass of the initial localized state translates in time while maintaining its shape for several transits of the ring.  An estimate of the magnitude of the rotation rate may be obtained as follows:  Assuming some portion of the center-of-mass OAM $\hbar p$ of the ground state is transferred to the localized state via the weak measurement, we find that $mNR|{ds\over dt}| \le \hbar |p|$, with ${ds\over dt}$ the velocity of the center-of-mass of the localized state, and the rotation rate obeys $|\Omega|={1\over R}|{ds\over dt}| \le {\hbar |p|\over mNR^2}$.  In our scaling the unit for time is $t_s = {2mR^2/\hbar}$, so the scaled rotation rate obeys $|v|=|\Omega|t_s \le {2|p|\over N}$, or $|v| \le 2$ for the current example.  Figure \ref{Fig4} shows an example of time-crystal-like behavior using $n=1$ and $\Delta\theta\sim 2.3$, which means a much weaker measurement than in Figure \ref{Fig3}(b). In this case the uncertainty is greater than the expected width $w\sim 1.6$ for a two-particle localized state, so the measurement is expected to be less disruptive.  This figure shows that the initial localized solution remains largely intact while it rotates with a velocity $v\sim 1$ \cite{Neq2}, and it persists over several rotations around the ring - an infinite number for the $N=2$ example since solutions of Eq. (\ref{EqNeq1}) must be periodic with period $2\pi$.  The initial position of the centroid of the localized solution is set by the outcome of the initial measurement, $\theta^{(0)}=0$ in Fig. \ref{Fig4}(b), and the centroid simply translates at velocity $v$ as time advances.  This example therefore exhibits time-crystal-like behavior.

The above argument may appear counter-intuitive since one would expect a localized initial state to expand and produce interference around the ring, giving rise to complicated dynamics. However, it is possible to construct solutions to the single-particle Schr\"odinger equation for a ring that are localised, non-dispersive, and rotating that have non-zero angular momentum \cite{Qin}.  Adapted to the solution of Eq. (\ref{EqNeq1}) these solutions take the form
\begin{equation}\label{rotst}
\psi(\theta_1,t) = {\cal N}_0\cdot\cos(m(\theta_1-v t))e^{i\ell\theta_1/2 + iv\theta_1/2 - i(m^2+v^2/4)t} ,
\end{equation}
where ${\cal N}_0$ is a normalization constant, and $m$ and $v$ are solution parameters. From Eq. (\ref{rotst}) we notice that $\psi(\theta_1,t)$ is a travelling localized solution with velocity $v$, and $2m$ is the number of wave function nodes around the ring.  For the simulation in Fig. \ref{Fig4} for $\ell=2$ there is one node around the ring, so $m=1/2$, and the velocity is $v=1$.  Moreover, one can verify that the corresponding initial wave function
\begin{equation}
\psi(\theta_1,t=0) = {\cal N}_0\cdot\cos\left ({\theta_1\over 2} \right )e^{3i\theta_1/2} ,
\end{equation}
is a physically allowed single-valued wave function that represents an excited state.  We have verified numerically that when the weak measurement is performed on the two-particle ground state using Eq. (\ref{psi_init}), the initialized wave function may be approximated near the origin by the above wave function $\psi(\theta_1,t=0)$.  The physical relevance of this is that, given our choice of measurement distribution in Eq. (\ref{G}) with $n=1$ and $\theta^{(0)}_2=0$, the weak measurement can dominantly excite a travelling wave solution of the ring that is initially centered at $\theta_1=0$ and that does not spread significantly over several transits of the ring.  In addition, because of the linearity of the single-particle Schr\"odinger equation (\ref{EqNeq1}), we can always construct superpositions of states such that we have a non-zero background as appears in the simulations.

\subsection{Role of the measurement strength}

The results shown in Figs. (\ref{Fig3}) and (\ref{Fig4}) show the extremes of strong and weak measurements, respectively, and how they impact the quantum dynamics as reflected in the probability density $p(\theta_1,t)$ of our system.  Next we provide some details of the transition between these extremes and how this may be quantified.

\begin{figure}[h!]
\centering
\includegraphics*[width=9cm]{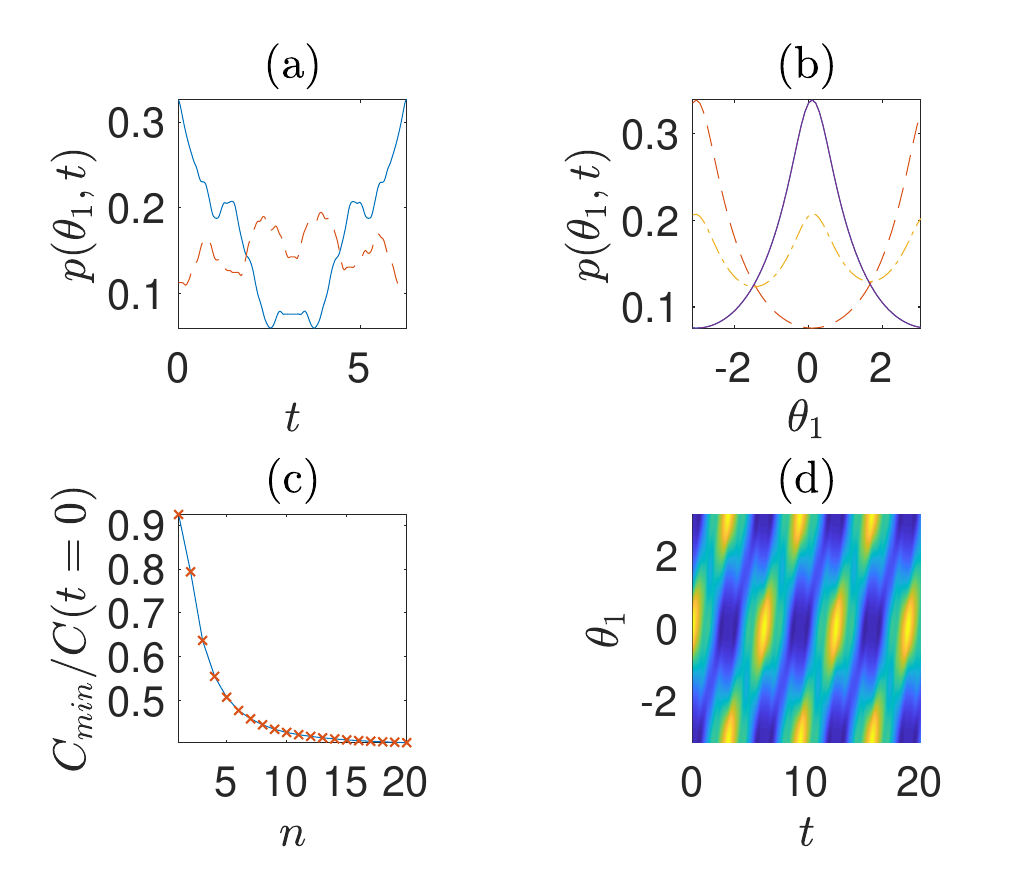}
\caption{(a) Probability density $p(\theta_1,t)$ versus $t=[0,2\pi]$ for $\theta_1=0$ (solid line) and $\theta_1=\pi/2$ (dashed line); (b) probability density versus $\theta_1=[-\pi,\pi]$ for $t=0,2\pi$ (solid line), $t=\pi/2$ (dash-dot line), and $t=\pi$ (dash line); (c) the normalized contrast $C_{min}/C(t=0)$ as a function of $n$; (d) Color coded plot of the probability density for $n=5$.  Parameters are the same as as in Fig. \ref{Fig1} along with $g=-2$.}\label{Fig5}
\end{figure}

To proceed we first consider the strong measurement example shown in Fig. \ref{Fig3}(b), and in Fig. \ref{Fig5}(a) we plot the corresponding probability density $p(\theta_1,t)$ versus $t=[0,2\pi]$ for $\theta_1=0$ (solid line) and $\theta_1=\pi/2$ (dashed line). Choosing $\theta_1=0$ the time variation of the probability density clearly displays periodic behavior with large contrast between the minimum and maximum values, whereas for $\theta_1=\pi/2$, while the periodicity is still evident, the contrast between the minimum and maximum values is reduced by a factor of two.  This dependence of the temporal variation of the probability density upon the choice of spatial angular position $\theta_1$, with concomitant variation in contrast, is a feature that arises when using a strong measurement for which $\Delta\theta/w<1$, that is the angular uncertainty $\Delta\theta$ of the measurement is less than the width $w$ of the $N=2$ solution.  The same conclusion arises when considering the spatial profile of the probability density for given times, and Fig. \ref{Fig5}(b) shows the probability density versus $\theta_1=[-\pi,\pi]$ for $t=0,2\pi$ (solid line), $t=\pi/2$ (dash-dot line), and $t=\pi$ (dash line).  The solid line shows both the initial localized probability density created by the measurement (plus the same superposed after $t=2\pi$), whereas at $t=\pi/2$ (dash-dot line) the probability density profile has significantly lower contrast.  The spatial profile of the probability density is also the same as the input at $t=\pi$, it is just shifted.  In contrast, for the example of a weak measurement with $\Delta\theta/w>1$ as illustrated in Fig. \ref{Fig4}, the contrast in the spatial probability density $p(\theta_1,t)$ for a given time remains constant to within a few percent.

The above discussion highlights that the contrast in the spatial profile of the probability density can serve as a quantitative measure of time-crystal-like behavior in our system.  Specifically, following SKS \cite{SKS_EPL} we here define the time dependent contrast as
\begin{equation}\label{C_t}
C(t) = { max_{\theta_1} [p(\theta_1,t)]-min_{\theta_1} [p(\theta_1,t)]\over max_{\theta_1} [p(\theta_1,t)]+min_{\theta_1} [p(\theta_1,t)]},
\end{equation}
and we expect time-crystal-like behavior to be absent when $C_{min} = min_t[C(t)]$ is quite a bit smaller than $C(0)$, whereas time-like behavior is possible when $C_{min} = min_t[C(t)] \simeq C(0)$.  Figure \ref{Fig5}(c) shows the normalized contrast $C_{min}/C(0)$ versus $n=[1,20]$ from our simulations, and we see that this has dropped to $0.5$ for $n=5$, for which $\Delta\theta/w=0.65$.  Consistent with this strong measurement we then find no time-crystal-like behavior for $n=5$ as illustrated in Fig. \ref{Fig5}(d) which shows the corresponding color-coded plot of the probability density $p(\theta_1,t)$ with concomitant degradation in the solution profile as it rotates in comparison to Fig. (\ref{Fig4}).

As noted earlier, for $N=2$ the quantum dynamics after the measurement must be periodic in time with period $2\pi$, so there is no true decay in this system.  On the other hand, following SKS \cite{SKS_EPL}, we can estimate the lifetime $t_c$ of the rotating solution as the minimum time for which the normalized contrast falls below $0.5$. Due to the periodicity any relevant lifetime must be less than $2\pi$.  Based on this definition we find from simulations that the lifetime for $n=50$ in Fig. \ref{Fig3}(b) is $t_c=1$, whereas for $n=5$ in Fig. \ref{Fig5}(d) it is $t_c = \pi/2$, both of which are shorter than the period $2\pi$ of one rotation.  In contrast, for $n=1$ the normalized contrast never falls below $0.9$, so the rotating solution remains largely intact for all times with some small superposed oscillations present, see Fig. \ref{Fig4}.  It is in this sense that our $N=2$ chiral soliton model can show time-crystal-like behavior to a good approximation.

The basic message from our two-particle simulations is that using a weak measurement of particle position time-crystal-like behavior is possible for the remaining particle on the ring if $\Delta\theta/w > 1$. This is the case since the weak measurement can dominantly excite a non-dispersive and rotating solution on the ring, whereas the more precise position measurement excites a broader wavepacket of the ring modes that leads to the observed loss of contrast and lack of  time-crystal-like behavior.  The benefit of the two-particle case is that it reveals the role played by the position measurement uncertainty in a transparent manner.

\section{Three-particle case}\label{3PC}

Although the two-particle case provides some insight it does not speak to the bigger issue of whether the spontaneously formed and rotating solution after the measurement can persist even for a multi-particle state, so the rotating solution may be viewed even approximately as a rotating soliton.  For this reason we next consider the three-particle case so that two particles are left on the ring after the measurement, with the possibility of exciting the two-particle analogue of a soliton \cite{Jackiw}.

\subsection{Three-particle Schr\"odinger equation}

The density for the $N=3$ bosonic particles may be written as
\begin{equation}
\rho(\theta_1,\theta_2,\theta_3) \equiv \eta(\theta_1-\theta_2)+\eta(\theta_1-\theta_3)+\eta(\theta_2-\theta_3) ,
\end{equation}
which by construction is symmetric under exchange of any pair of particle coordinates.  The Schr\"odinger equation for $\Psi(\theta_1,\theta_2,\theta_3,t)$ may then be written as
\begin{eqnarray}
i{\partial\Psi\over\partial t} &=& \sum_{j=1}^3  \left ( -i{\partial\over \partial \theta_j}- {\ell\over 2} -\kappa\rho(\theta_1,\theta_2,\theta_3)) \right )^2 \Psi \nonumber \\ && + {g\over 2} \rho(\theta_1,\theta_2,\theta_3) \Psi .
\end{eqnarray}
To proceed it is useful to use the following Jacobi coordinates \cite{Jacobi} appropriate to this three-body problem
\begin{equation}
s = {1\over 3} (\theta_1+\theta_2+\theta_3), \quad
x = (\theta_1-\theta_2), \quad
y = (\theta_1+\theta_2-2\theta_3),
\end{equation}
and inverting the Jacobi coordinates we obtain
\begin{eqnarray}
\theta_1 &=& s + {1\over 6}y + {1\over 2}x, \nonumber \\
\theta_2 &=& s + {1\over 6}y - {1\over 2}x  , \nonumber \\
\theta_3 &=& s - {1\over 3}y. 
\end{eqnarray}
From these results we find
\begin{eqnarray}
(\theta_1-\theta_2) &=& x \nonumber \\ 
(\theta_1-\theta_3) &=& {1\over 2}y + {1\over 2}x, \nonumber \\
(\theta_2-\theta_3) &=& {1\over 2}y - {1\over 2}x  ,
\end{eqnarray}
so that the density $\rho$ in these Jacobi coordinates becomes
\begin{equation}
\rho(x,y) = \eta \left ({x\over 2} \right ) + \eta \left ( {y\over 4} + {x\over 4} \right ) + \eta \left ( {y\over 4} - {x\over 4} \right ),
\end{equation}
which is independent of the COM coordinate $s$.  The transformed Schr\"odinger equation for $\Psi(x,y,s,t)$ then becomes
\begin{eqnarray}\label{Psi3}
i{\partial\Psi\over\partial t} &=& -\left ( 2{\partial^2\over \partial x^2} + 6{\partial^2\over \partial y^2} + {1\over 3} {\partial^2\over \partial s^2} \right )\Psi + 2i\kappa\rho(x,y){\partial\Psi\over\partial s}  \nonumber \\ && + 3\left [{\ell\over 2} + \kappa\rho(x,y) \right ]^2\Psi + {g\over 2}\rho(x,y)\Psi .
\end{eqnarray}

\subsection{Three-particle ground state}

We seek a three-particle stationary state using the ansatz
\begin{equation}
\Psi(x,y,s,t) = e^{-i\varepsilon t + ips}\varphi (x,y),
\end{equation}
where $p$ is the integer valued winding number associated with the center-of-mass OAM of the three particles.  Substituting the above ansatz into Eq. (\ref{Psi3}) yields
\begin{eqnarray}\label{phi3Eq}
&&\varepsilon \varphi(x,y)  = -\bigg ( 2{\partial^2\over \partial x^2} + 6{\partial^2\over \partial y^2} \bigg)\varphi \nonumber \\ && + 3\bigg ( {p\over 3} -{\ell\over 2} - \kappa\rho(x,y) \bigg)^2\varphi + {g\over 2}\rho(x,y)\varphi ,
\end{eqnarray}
where $\varepsilon$ is the scaled energy of the three-particle system. 

\begin{figure}[h!]
\centering
\includegraphics*[width=6.2cm]{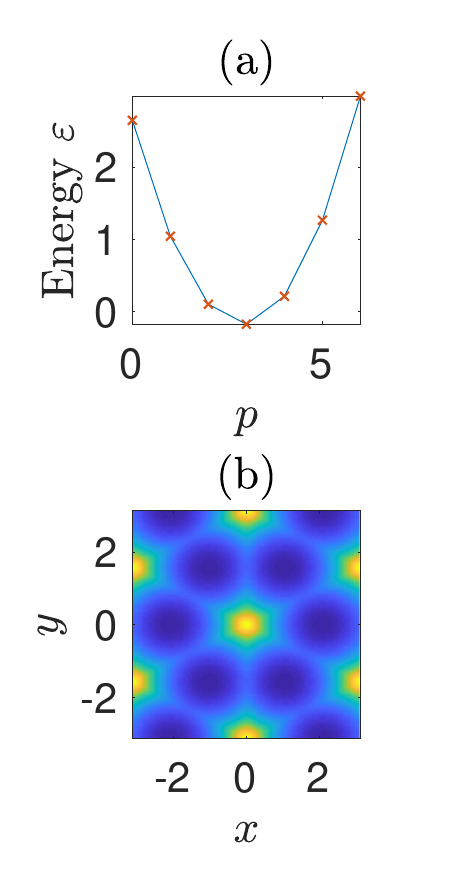}
\caption{(a) Scaled energy $\varepsilon$ versus COM winding number $p$, and (b) ground state profile $|\varphi(x,y)|^2$ for $p=3$.  Parameters are $\ell=2,g=-0.7,\kappa=0.05, q=50$.}\label{Fig6}
\end{figure}

We have solved Eq. (\ref{phi3Eq}) numerically for a range of parameters using the imaginary time method \cite{LehToiElo2007}, in particular to obtain the ground state $\varphi(x,y)$ and associated scaled energy $\varepsilon$.  Figure \ref{Fig6} shows numerical results for parameters $\ell=2,g=-0.7,\kappa=0.05, q=50$.  Specifically, Fig. \ref{Fig6}(a) shows the scaled energy versus $p$ and reveals that the ground state occurs when ${p\over N} = {\ell\over 2}$, or $p=3$ for this example, and (b) shows a line plot of the ground state profile $|\varphi(x,y)|^2$ versus $x$ and $y$ for $p=3$.

\subsection{Time-crystal-like dynamics from measurement}

\begin{figure}[h!]
\centering
\includegraphics*[width=5cm]{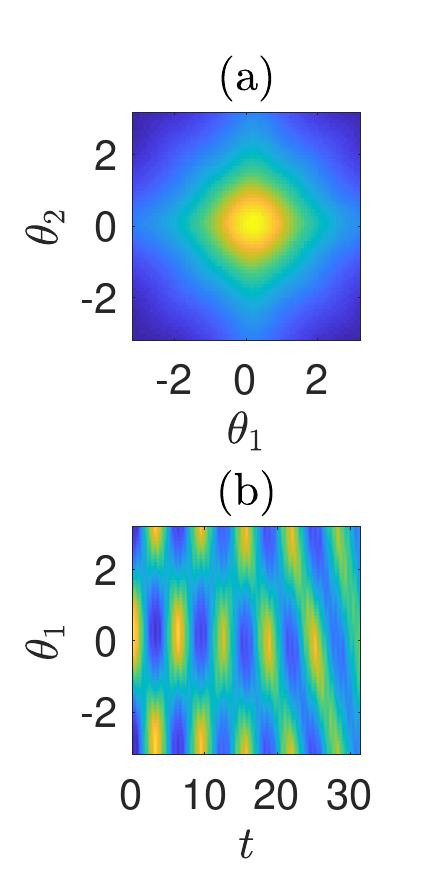}
\caption{(a) Reconstructed ground state probability density $|\Psi(\theta_1,\theta_2,\theta_3=0)|^2$, and (b) simulation of the single-particle probability density $p(\theta_1,t)$ for a measurement with $n=50$ that localizes particle $3$ with an uncertainty $\Delta\theta\sim 0.34$.  This example has $\Delta\theta/w = 0.24$ and does not display time-crystal-like behavior.  Parameters are the same as as in Fig. \ref{Fig6}.}\label{Fig7}
\end{figure}

To proceed we first recognize that the ground state represented by the probability density in Fig. \ref{Fig6}(b) does not correspond to a localized state on the ring since $|\varphi(x,y)|^2$ depends only on the Jacobi coordinates $x$ and $y$, and there is no information concerning the COM position $s$ of the particle. Again, since the ground state has a winding number $p\ne 0$ it does represents a persistent flow \cite{RyuAndCla07}. We can also reconstruct the ground state three-body wave function for one coordinate fixed, say $\theta_3=0$, and $|\Psi(\theta_1,\theta_2,\theta_3=0)|^2$ (at $t=0$) is shown in Fig. \ref{Fig7}(a). We note that the width (full-width at half-maximum) of this ground state density profile is $w\sim 1.4$, that provides a characteristic scale for localized three-particle states that may appear in the system.

A few words are in order about transforming from $\varphi(x,y)$ to $\Psi(\theta_1,\theta_2,\theta_3)$.  First we note that the Jacobi coordinates $x$ and $y$ can be re-expressed as
\begin{equation}
    x= \theta_1' - \theta_2', \quad y = \theta_1' + \theta_2',
\end{equation}
where $\theta_{1,2}' = \theta_{1,2}-\theta_3$.  This means that, for a given $\theta_3$, the transformation between the $\varphi(x,y)$ in the Jacobi coordinates $x$ and $y$, and $\Psi(\theta_1',\theta_2')$ in the coordinates $\theta_1'$ and $\theta_2'$ can be performed using a $45^\circ$ rotation.  We note that at this point the wave function $\Psi(\theta_1',\theta_2')$ should be symmetrized if needed.  One can then transform from $\Psi(\theta_1',\theta_2')$ to $\Psi(\theta_1,\theta_2,\theta_3)$ using simple linear displacements $\theta_{1,2} = \theta_{1,2}'+\theta_3$.

We next extend the measurement model to the case with $N=3$ particles.  In particular, we assume that the position of particle $3$ is measured yielding a value $\theta_3^{(0)}$, but allow for the measurement to be imprecise, so that the wave function for the remaining particles may be written as
\begin{eqnarray}\label{psi_3init}
  &&  \psi(\theta_1,\theta_2,t=0)=\Psi(\theta_1,\theta_2,\theta_3^{(0)}) =\nonumber \\ && \int_{-\pi}^\pi d\theta_3~G(\theta_3-\theta_3^{(0)})\Psi(\theta_1,\theta_2,\theta_3) ,
\end{eqnarray}
where the function $G(\theta_3-\theta_3^{(0)})$ reflects the uncertainty of the position measurement, and we again use the model in Eq. (\ref{G}). After the measurement the wave function $\psi(\theta_1,\theta_2,t=0)$ (suitably normalized) for the remaining particles is given by Eq. (\ref{psi_3init}), and subsequent time development is governed by the two-particle Schr\"odinger equation
\begin{eqnarray}\label{EqNeq2}
i{\partial\psi\over\partial t} &=& -\bigg( {\partial^2\over\partial \theta_1^2} +  {\partial^2\over\partial \theta_2^2} \bigg)\psi 
\nonumber \\ &&
+2i \bigg( {\ell\over 2} +\kappa\rho \bigg )\bigg({\partial\over\partial \theta_1} + {\partial\over\partial \theta_2} \bigg)\psi 
\nonumber \\ &&
+ 2 \bigg( {\ell\over 2}  +\kappa\rho \bigg)^2\psi + {g\over 2}\rho \psi, 
\end{eqnarray}
where $\rho=\eta(\theta_1-\theta_2)$ is the scaled density for the two remaining particles.

Figure \ref{Fig7}(b) shows a simulation of the quantum dynamics according to Eq. (\ref{EqNeq2}) following a precise measurement for which we set $\theta_3^{(0)}=0$ without loss of generality, and $n=50$ so that $\Delta\theta\sim 0.34$ (full-width at half-maximum of $|G(\theta)|^2$): As for the $N=2$ case, we chose $n=50$ as this approaches a precise measurement as closely as possible within our non-local model with $q=50$ in Eq. (\ref{G}). In particular, Fig. \ref{Fig7}(b) shows a color-coded plot of the single-particle probability density for measuring a second particle
\begin{equation}
    p(\theta_1,t) = \int_{-\pi}^\pi d\theta_2 |\psi(\theta_1,\theta_2,t)|^2 .
\end{equation}
In this case the uncertainty is much less than the expected width $w\sim 1.4$ for a three-particle localized state, so the measurement is expected to be very disruptive.  We note that although the probability density appears to be varying periodic in time with period $2\pi$, for $t>4\pi$ the profile is seen to deviate significantly from this periodicity.  Moreover, as expected for a strong measurement, the initial localized solution evident in the single-particle probability density $p(\theta_1,t=0)$ becomes rapidly distorted as time increases, not simply displaced as expected for a time crystal, so this example does not display time-crystal-like behavior.  This is in perfect keeping with the findings of SKS that too precise a position measurement of one particle can produce a localized state initially but this will become rapidly dispersed before the initial localized solution can exhibit even a single rotation around the ring.

\begin{figure}[h!]
\centering
\includegraphics*[width=8cm]{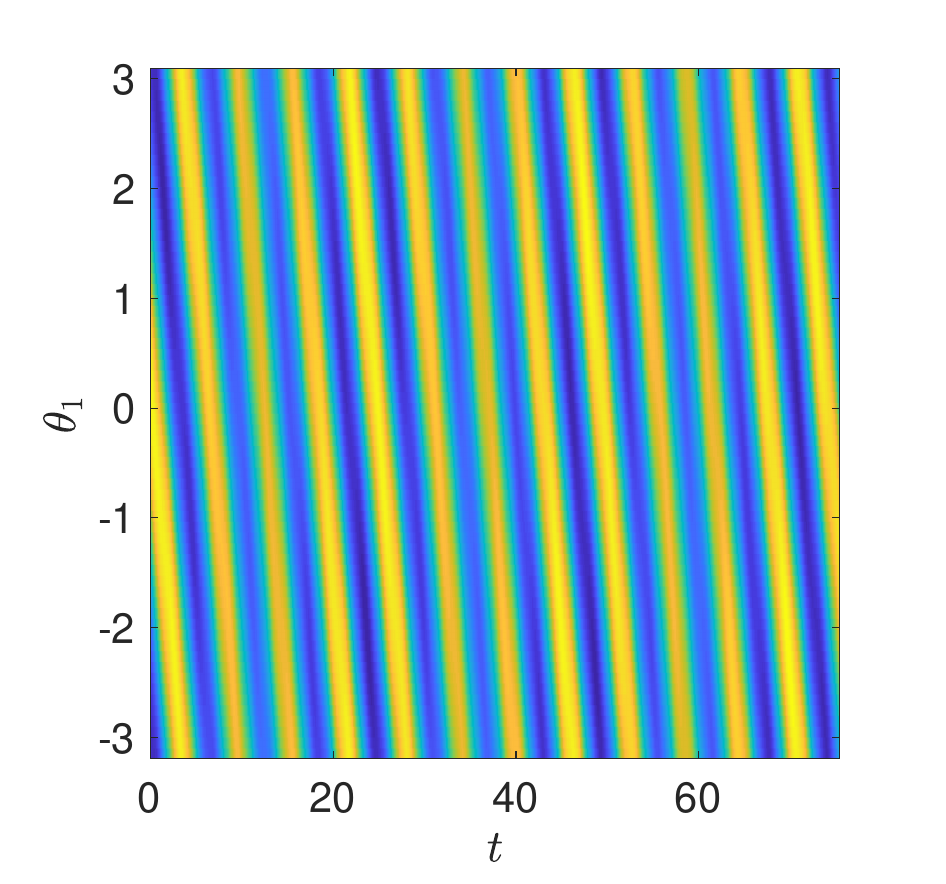}
\caption{Simulation of the single-particle probability density $p(\theta_1,t)$ for an imprecise measurement with $n=1$ and an uncertainty $\Delta\theta\sim 2.3$ that couples the remaining two particles after the measurement of particle $3$ to the two-particle limit of the chiral soliton.  This example has $\Delta\theta/w = 1.63$ and does display time-crystal-like behavior. Parameters are the same as as in Fig. \ref{Fig6}.}\label{Fig8}
\end{figure}

For time-crystal-like behavior what is required is that the center of the initial localized state basically translates in time and persists in shape for several rotations around the ring. We may use our previous estimate for the magnitude of the scaled rotation rate $|v| \le {2|p|\over N}$, or $|v|\le 2$ for the current example.  Figure \ref{Fig8} shows an example of  time-crystal-like behavior using $n=1$ and $\Delta\theta\sim 2.3$, which means a much weaker measurement than in Figure \ref{Fig7}(b). In this case the uncertainty is greater than the expected width $w\sim 1.4$ for a three-particle localized state, so the measurement is expected to be less disruptive.  This figure shows that the initial localized solution remains largely intact while it rotates with velocity $v\sim -1$ \cite{Neq3}, and in the figure is seen to persist over $> 10$ rotations around the ring.  This example therefore exhibits time-crystal-like behavior to a good approximation.

To summarise, we deduce that the initial measurement of particle $3$ leads to spontaneous formation of a localized state that is a two-particle analogue of a mean-field chiral soliton \cite{Jackiw}, and that this two-particle soliton can continue rotating largely intact over several rotation periods.  This can happen here since we considered a weak measurement as opposed to the precise measurement that leads to decay of the rotating soliton, see Fig. \ref{Fig7}(b).

\subsection{Role of the measurement strength}

The results shown in Figs. (\ref{Fig7}) and (\ref{Fig8}) show the extremes of strong and weak measurements, respectively, and how they impact the quantum dynamics for $N=3$ particles as reflected in the probability density $p(\theta_1,t)$ of the system.  We next provide some details of how the strength of the measurement impacts whether time-crystal-like behavior is present or not.

\begin{table}[ht]
\caption{Data on measurement strength}\label{Table1}
\begin{center}
\begin{tabular}{|c|c|c|} \hline
n & $\Delta\theta/w$ & $t_c/2\pi$ \\ \hline\hline
1 & 1.63 & $> 50$ \\ \hline
2 & 1.16 & 1.7 \\ \hline
3 & 0.96 & 0.7 \\ \hline
4 & 0.83 & 0.2 \\ \hline
5 & 0.74 & 0.2 \\ \hline
6 & 0.68 & 0.2 \\ \hline
\end{tabular}
\end{center}
\end{table}

Turning first to the case with a strong measurement in Fig. \ref{Fig7}(b) with $n=50$ and $\Delta\theta/w = 0.24$, we note that the single-particle probability density $p(\theta_1,t)$ is no longer periodic in time. This is the case since the quantum dynamics after the measurement is now governed by the two-dimensional Schr\"odinger equation (\ref{EqNeq2}) which will have a more complex eigenmode spectrum than the linear Eq. (\ref{EqNeq1}) for the case of $N=2$ particles.  As a result the wave packet dynamics for the case of $N=3$ particles can be far richer.  In particular, the contrast $C(t)$ defined in Eq. (\ref{C_t}) can now get progressively smaller as time increases, though recurrences may still occur in general.  Indeed, SKS \cite{SKS_EPL} defined the lifetime $t_c$ of the revolving solution as the first time for which the normalized contrast $C(t_c)/C(0)=0.5$ drops to one-half its initial value.  We adopt this approach here, and for the example in Fig. \ref{Fig7}(b) $t_c \simeq 0.2$ meaning that the initial localized solution executes only a fraction of a revolution of the ring that takes time $2\pi$.  In contrast, in Fig. \ref{Fig8} with $n=1$ and $\Delta\theta/w = 1.63$, time-crystal-like behavior is clear over the duration $t=[0,80]$ of the simulation, which is greater than $t_c/2\pi >10$ revolutions of the ring, and the normalized contrast remains $>0.9$ throughout. We limited the maximum time to $80$ only so the features in Fig. \ref{Fig8} were resolvable.  Indeed, for times over which we can maintain numerical accuracy, which amounts to more than $t_c/2\pi> 50$ revolutions, the normalized contrast remains well above $0.5$.  Within numerical accuracy it is therefore legitimate to claim that this $N=3$ example clearly shows time-crystal-like behavior, though we cannot claim that it persists indefinitely.

To explore how the time-crystal-like behavior varies with the strength of the measurement we performed simulations for a range of different values of $n$ in Eq, (\ref{G}), and some results are tabulated in Table \ref{Table1}.  The left column gives $n$, the center column gives the corresponding value for $\Delta\theta/w$, and the right hand column gives $t_c/2\pi$ which provides an estimate of the number of revolutions executed by the solution before it decays. We see that for $n \ge 4$ the solution executes only a fraction of a revolution, which is consistent with the fact that these correspond to strong measurements with $\Delta\theta/w < 1$.  In contrast, for $n=2,3$ for which $\Delta\theta/w \simeq 1$, the solution executes roughly $1-2$ revolutions before decaying.  Finally, for $n=1$ and $\Delta\theta/w=1.63$ we have the example shown in Fig. \ref{Fig8} where the lifetime increases dramatically to $>50$, indicative of a divergence.  This apparent divergence in the lifetime signals that the time-crystal-like behavior can persist for a large number of revolutions of the ring.

\section{Summary and conclusions}\label{SC}

In 2019 we proposed a mean-field chiral soliton model for a quantum time crystal \cite{OW19}, and this proposal met with some criticism in the literature \cite{SKS_EPL},\cite{SKS_comment}-\cite{OW_lack}.  The goal of the present paper was to examine a few boson $(N=2,3)$ limit of our previous chiral soliton model to numerically assess whether time-crystal-like behavior is possible in the context of the type of calculation in Ref. \cite{SKS_EPL}.  In particular, we initialize the system in the $N$-boson ground state, perform a position measurement of one particle, and see if spontaneous formation of the resulting $(N-1)$-boson localized state (soliton for $N=3$)  can persist over several revolutions of the ring as required for time-crystal-like behavior.  We find that for an imprecise or weak position measurements time-crystal-like behavior is possible in our chiral soliton model, whereas for a precise measurement quantum fluctuations cause the soliton to decay in accordance with \cite{SKS_EPL}.

One feature from the simulations is that in the presence of chirality the quantum ground state of the system can carry orbital angular momentum $\hbar p$, with $p$ the center-of-mass winding number. This means that the lowest energy state can be a persistent flow \cite{RyuAndCla07}. The point is that the presence of chirality and a ground state that is a persistent flow appear to be key requirements for time-crystal behavior to be a possibility in our model, the imprecise measurement facilitating transfer of the orbital angular momentum to the center-of-mass momentum of the localized soliton.  Estimates for the rotation rates were given but details of this process remain to be elucidated.

\begin{figure}[h!]
\centering
\includegraphics*[width=8cm]{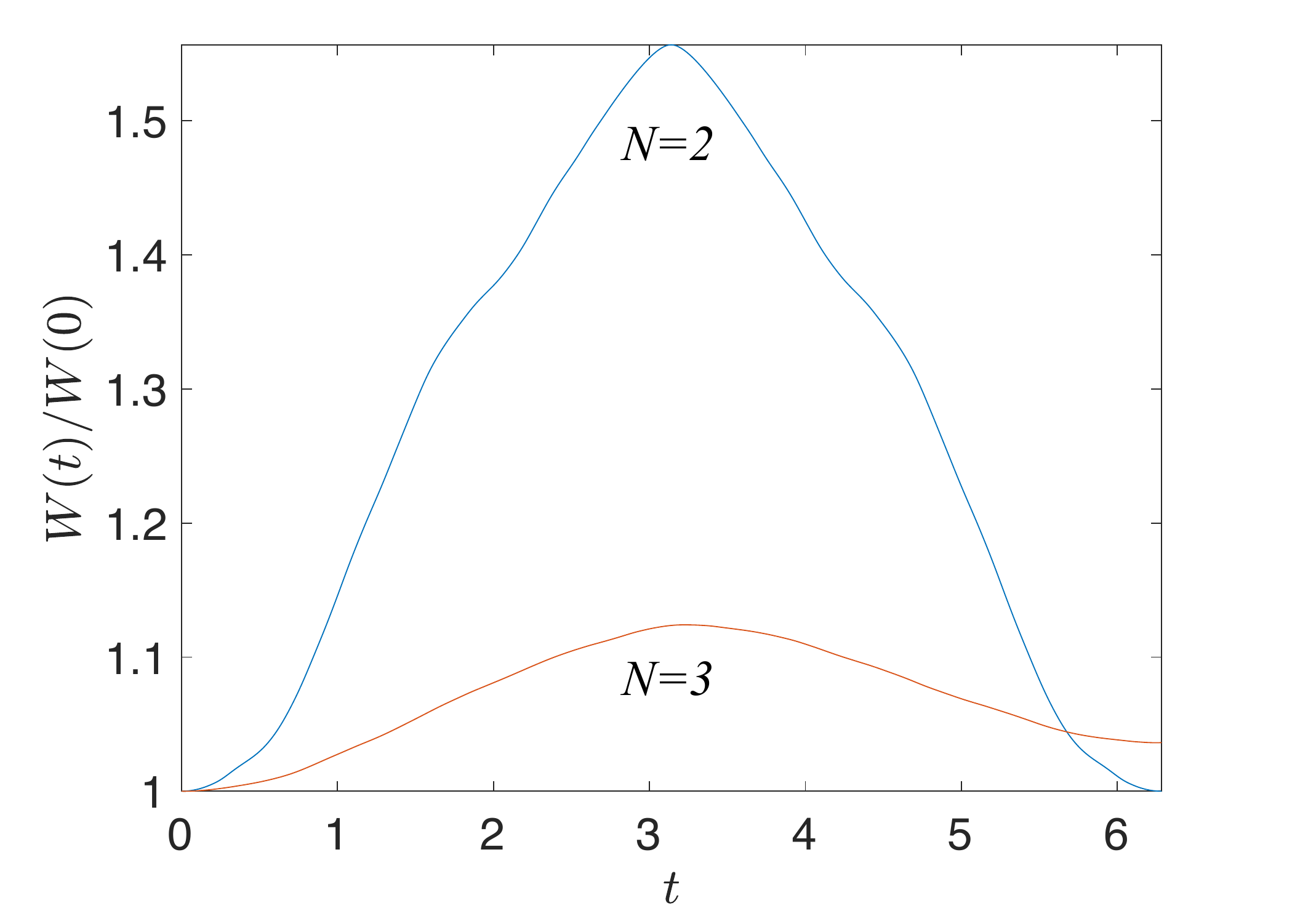}
\caption{The plot shows the normalized root-mean-square width $W(t)/W(0)$ after a weak measurement $(n=1)$ versus $t$ for $N=2$ and $N=3$ for parameters $\kappa=0.2, g=-2.0$.}
\label{Fig9}
\end{figure}

Clearly the small particle numbers and parameters used here are not compatible with experiments, but are rather intended to show that time-crystal-like behavior is not ruled out in our model.  We remark that the same behavior could be seen for other combinations of parameters as long as a weak measurement is employed, with $\Delta\theta/w > 1$.  Restricting to a few bosons allowed for numerical simulations to be performed for our chiral soliton model without undue approximations, but our conclusions may still be of relevance in the more general case with large $N$. According to the Wilczek model analyzed in Ref. \cite{SKS_EPL} the soliton lifetime $t_c$ is expected to scale with $N$.  If this scaling also applies to our chiral model, then if $N=3$ can show time-crystal-like behavior it should be expected for larger $N$ as the lifetime increases.  To explore the scaling of the lifetime with $N$ we calculated the root-mean-square width $W(t)$ of the evolving solution after a weak measurement $(n=1)$ over the time range $t=[0,2\pi]$, and Fig. \ref{Fig9} shows $W(t)/W(0)$ for $N=2$ and $N=3$ for parameters $\kappa=0.2, g=-2.0$ (similar results were found for the parameters $\kappa=0.05, g=-0.7$).  In particular, the plot shows that the width $W(t)$ initially grows more slowly for the case $N=3$ compared to $N=2$, and this adds some credibility to the statement that “the soliton lifetime $t_c$ is expected to scale with $N$.”  The reversal of the initial increase in $W(t)$ is a consequence of the recurrence that occurs due to the periodic boundary conditions imposed by the ring, this leading to the fact that for $N=2$ the solution must be strictly periodic in time with period $2\pi$.  In future publications we plan to extend this initial numerical study to larger particle numbers, both numerically  and analytically, and thereby move the project towards experimentally accessible parameter ranges.

\section{Acknowledgements}

We acknowledge helpful discussions with Brian Anderson, Erika Andersson, Stewart Lang, Joel Priestley, and Gerard Valent\'{\i}-Rojas.


\begin{thebibliography}{99}
\bibitem{Neq2}
To obtain the sign of the rotation rate requires not only the sign of the ground state winding number $p$ but also the sign of the localized state energy.  For the current example with $N=2$ we have $p=2$ and the energy of the single-particle localized state is positive, it is localized but not a bound state.  In this example the rotation rate is positive.
\bibitem{Neq3}
For the current example with $N=3$ we have $p=3$ but the energy of the two-particle localized soliton state is negative by virtue of being a bound state.  In this case the rotation rate is negative.
\end{thebibliography}
\end{document}